\documentclass[10pt,aps,prc,twocolumn,showpacs,showkeys,amsmath,floatfix,superscriptaddress,preprintnumbers]{revtex4-1}
\usepackage{color,graphicx}
\usepackage{mathptmx}                
\usepackage{dcolumn}                 
\usepackage{bm}                      
\usepackage{ulem,soul}
\usepackage{nicefrac}

\usepackage[dvipsnames,usenames]{xcolor}

\begin{document}

\preprint{INT-PUB-18-028}

\title{On the effect of high order empirical parameters on the nuclear equation of state}

\author{J\'er\^ome Margueron}
\affiliation{Institute for Nuclear Theory, University of Washington, Seattle, Washington 98195, USA}
\affiliation{Institut de Physique Nucl\'eaire de Lyon, CNRS/IN2P3, Universit\'e de Lyon, Universit\'e Claude Bernard Lyon 1, F-69622 Villeurbanne Cedex, France}

\author{Francesca Gulminelli}
\affiliation{CNRS, ENSICAEN, UMR6534, LPC ,F-14050 Caen cedex, France}

\date{\today}

\begin{abstract}
A quantitative knowledge of the nuclear Equation of State (EoS) requires an accurate estimation of the uncertainties on the  EoS parameters
and their mutual correlations. Such correlations are empirically observed in a large set of EoS models by different authors, but they are not always fully understood. We show that some of these correlations can be interpreted from basic physical constraints  imposed on a simple Taylor expansion of the binding energy  around saturation density. 
In particular, we investigate the correlations among the following empirical parameters: the symmetry energy $E_{sym}$, the slope and curvature of the symmetry energy $L_{sym}$ and $K_{sym}$, and the curvature and skewness of the binding energy in symmetric matter $K_{sat}$ and $Q_{sat}$. 
The uncertainties on these correlations is estimated through analytical modelling as well as a meta-modelling analysis of the EoS subject to physical constraints. We show that a huge dispersion of the correlations among low order empirical parameters is induced by the unknown higher order empirical parameters, such as $K_{sym}$ (second order) and $Q_{sat}$ and $Q_{sym}$ (third order).
We also propose an explanation of the reason why $Q_{sat}$ is weakly constrained by present experimental data.
We conclude that selective observables on high order parameters, such as $K_{sym}$ and $Q_{sat}$, should be better determined before the present uncertainties on the EoS can be further reduced.
\end{abstract}

\maketitle

\section{Introduction}

The nuclear Equation of State (EoS) around the saturation density of symmetric matter ($n_{sat}$) can be accurately characterized by the so-called empirical parameters, defined as the set of successive derivatives of the energy functional~\cite{Myers1969}. Since the density of atomic nuclei is not far from $n_{sat}$, nuclear experiments essentially probe the low order empirical parameters, e.g. up to order two or so. The better known the empirical parameters, the more precise the EoS around $n_{sat}$. With the continuous improvement of theoretical modeling and of the nuclear data, the knowledge of these  parameters as well as of the correlations among them have considerably progressed in the recent years, inducing tighter constraints on the nuclear EoS. Typical examples are given by the determination of the saturation energy from mass measurements~\cite{Goriely2015,Steiner2015,Bertsch2017,McDonnel2015}, the nuclear incompressibility from the Giant Monopole Resonance (GMR)~\cite{Blaizot1980,Colo2004,Khan2012,Khan2013,Colo2014}, and the symmetry energy properties ($E_{sym}$ and $L_{sym}$) from various experiments such as mass measurements~\cite{Kortelainen2012,Danielewicz2014}, isovector giant resonances~\cite{Trippa2008,Colo2014}, the correlation between $L_{sym}$ and the surface stiffness parameter for the determination of the neutron skin~\cite{Warda2009,Centelles2010,Mondal2016}, or more recently the constraints on $K_{sym}$ induced by the unitary limit in neutron matter~\cite{Tews2017}. 
These constraints are summarized in various reviews, such as Refs.~\cite{Lattimer2013,Newton2014,BALi2014} for instance.

Most of the experimental determination of the empirical parameters so far relies on the assumed linear correlation between an experimental observable and a single empirical parameter. For such analyses, the final uncertainty on each empirical parameter -- and therefore on the global EoS -- crucially depends on the quality of the correlation. In addition, the quality of the correlation depends on the variation range of the other parameters of the set, which can be model dependent. A way to settle the model dependence is to look for specific correlations between empirical parameters which would be induced by the specific functional form of the chosen model. This was for instance shown to be the case for the experimental determination of $K_{sat}$. The correlation between $K_{sat}$ and $Q_{sat}$ typically found for Skyrme and Gogny interactions, is related to the presence of a single density dependent term in the nuclear force~\cite{Khan2012,Khan2013}. The density dependence of the energy per particle in symmetric matter being controlled by $K_{sat}$ and $Q_{sat}$ at first orders, it is important to understand such model dependence, as discussed in Ref.~\cite{Margueron2018a,Margueron2018b}.

Since experimental measurements are often sensitive to both the bulk and surface properties of finite nuclei, they probe a continuous range in density, implying that there is no one-to-one correspondence between observables and EoS parameters. Each experimental probe is sensitive to a set of parameters, possibly leading to some model dependence in the determination of single parameters.  The determination of the EoS  thus comes through the intersection between the different correlation plots among empirical parameters, as established through the comparison of density functional calculations to the different observables~\cite{Tsang2012,Lattimer2013,Lattimer2014,Dutra2014,Fortin2016}. 
In the absence of a global analysis involving most of the possible model dependence, the question of the EoS uncertainties remains however unsolved.

To progress on the question of the model dependence, a meta-modeling was proposed~\cite{Margueron2018a}, where the variation of the empirical parameters is set to be free and only constrained by the physical requirements imposed to the meta-modeling, e.g., existence of the saturation point, stability of the EoS, positiveness of the symmetry energy, causality, constraints given by ab-initio calculations at low density, etc.... Based on this global analysis satisfying a set of physical requirements, generic correlations among low-order empirical parameters have been analyzed, see for instance Refs.~\cite{Margueron2018b}. These correlations are usually found to be weaker than the ones deduced using both a specific model and a direct fit to nuclear properties, e.g. Skyrme or relativistic mean field, see Refs.~\cite{Kortelainen2012,Danielewicz2014}.

In this paper, we focus on simple cases where the correlations among empirical parameters can be analyzed from general nuclear matter properties. We try to estimate how much the correlations between low order empirical parameters are blurred by the uncertainties on higher-order ones. To this aim, we estimate the propagation of the high-order parameter uncertainties down to the lower order ones, based on a simple Taylor expansion of the EoS around the saturation density $n_{sat}$, presented in Sec.~\ref{sec:def}. In Sec.~\ref{sec:results} the quality of these error estimations is then checked against the prediction of a set of $\sim$50 different realistic nuclear functionals. Finally, in Sec.~\ref{sec:MM} a more complete analysis of the correlations is performed within a meta-model of the equation of state~\cite{Margueron2018a}, in which several hundreds of thousand different functionals are generated assuming full independence among the empirical parameters, and subsequently filtered through many-body perturbation theory (MBPT) predictions based on chiral effective field theory ($\chi$EFT) interactions, stability of the EoS and causality conditions.

\section{Definitions and strategy}
\label{sec:def}

Given a generic functional for the energy per particle of homogeneous nuclear matter $e(n_n,n_p)$, simply expressed as the sum of an isoscalar $e_{sat}(n)$ and isovector $e_{sym}(n)$ terms ($n=n_n+n_p$, $\delta=(n_n-n_p)/n$),
\begin{equation}
e(n_n,n_p) = e_{sat}(n) + e_{sym}(n) \delta^2 + ...,
\label{eq:energy}
\end{equation}
where the small contribution from non-quadratic terms are neglected here, the isoscalar empirical parameters are defined as
the successive density derivatives of $e_{sat}(n)$, 
\begin{equation}
P_{IS}^{(k)}= (3n_{sat})^k \frac{\partial^k e_{sat}}{\partial n^k}|_{\delta=0,n=n_{sat}}, 
\label{eq:empIS}
\end{equation}
We will note $P_{IS}^{(0)}=E_{sat}$ the saturation energy, 
$P_{IS}^{(2)}=K_{sat}$ the incompressibility, $P_{IS}^{(3)}=Q_{sat}$ the skewness, and $P_{IS}^{(4)}=Z_{sat}$ the kurtosis.
In Eq.~(\ref{eq:energy}), $e_{sym}(n)$ is the symmetry energy function of the density $n$ and defined as $e_{sym}=\nicefrac{1}{2}\partial^2e/\partial\delta^2|_{\delta=0}$ in symmetric matter.
The isovector parameters measure the density derivatives of the symmetry energy as,
\begin{equation}
P_{IV}^{(k)}=  (3n_{sat})^k \frac{\partial^k e_{sym}}{\partial n^k}|_{\delta=0,n=n_{sat}} .
\label{eq:empIV}
\end{equation}
We will note $P_{IV}^{(0)}=E_{sym}$ the symmetry  energy at saturation,  $P_{IV}^{(1)}=L_{sym}$ the symmetry energy slope,
$P_{IV}^{(2)}=K_{sym}$ the isovector incompressibility, $P_{IV}^{(3)}=Q_{sym}$ the isovector skewness, and $P_{IV}^{(4)}=Z_{sym}$ the isovector kurtosis.

A Taylor expansion around the saturation density $n_{sat}$ is naturally suggested by the definition of the empirical parameters  (\ref{eq:empIS})-(\ref{eq:empIV}), and depending on the truncation of the Taylor series we will have different approximations for the functional $e(n_n,n_p)$ as:
\begin{eqnarray}
e_{sat,2}(x)&=&E_{sat}+\frac 1 2 K_{sat} x^2  \, , \label{eq:esat2}\\
e_{sat,3}(x)&=&E_{sat}+\frac 1 2 K_{sat} x^2 + \frac 1 6 Q_{sat} x^3   \, , \\
e_{sat,4}(x)&=&E_{sat}+\frac 1 2 K_{sat} x^2 + \frac 1 6 Q_{sat} x^3 + \frac 1 {24} Z_{sat} x^4   \, , \label{eq:esat4}
\end{eqnarray}
and
\begin{eqnarray}
e_{sym,2}(x)&=&E_{sym}+L_{sym} x + \frac 1 2 K_{sym} x^2  \, ,  \label{eq:esym2}\\
e_{sym,3}(x)&=&E_{sym}+L_{sym} x + \frac 1 2 K_{sym} x^2 + \frac 1 6 Q_{sym} x^3   \, , \\
e_{sym,4}(x)&=&E_{sym}+L_{sym} x + \frac 1 2 K_{sym} x^2 + \frac 1 6 Q_{sym} x^3 
+ \frac 1 {24} Z_{sym} x^4   \, . \nonumber \\ \label{eq:esym4}
\end{eqnarray}
where the parameter $x$ is introduced for convenience and is defined as $x=(n-n_{sat})/(3n_{sat})$. The empirical parameters (\ref{eq:empIS})-(\ref{eq:empIV})
can be identified as the coefficients of the expansion in Eqs.~(\ref{eq:esat2})-(\ref{eq:esym4}), where we adopt the naming usage for the empirical parameters.
Note however that the convention may depend on the authors, see the appendix of Ref.~\cite{Piekarewicz2009} for a detailed discussion. 

In principle, both the isospin expansion (\ref{eq:energy}) and the density expansions (\ref{eq:empIS})-(\ref{eq:empIV}) could be performed beyond the orders we considered here. For the characterization of the nuclear EoS between 0 and $n_{sat}$ that we analyze here, the proposed expansions (\ref{eq:esat2})-(\ref{eq:esym4}) are found to be sufficient.

From the series expansion of the functional $e(n_n,n_p)$, it is clear that any direct measurement or physical constraint on the functional will naturally produce some correlations among the empirical parameters $P^{(k)}$. Let us consider, for instance, an observable $\langle O(x)\rangle=f(e_{sym})$ that we suppose to be both sensitive to the isovector part of the functional and independent of the terms $\propto x^k$, $k\geq 2$, where this last condition will be met if, e.g. the observable is defined sufficiently close to saturation density. The constraint of reproducing the observable $\langle O\rangle$ would naturally produce an exact linear correlation between the parameters $E_{sym}$ and $L_{sym}$. However, in a realistic application, the higher order terms $k\geq 2$ are never fully negligible, and might blur such correlation. 

In the following sections, we will work out the different correlations among empirical parameters implied in Eqs.~(\ref{eq:esat2})-(\ref{eq:esym4}),
when the value of the energy functional $e(n_n,n_p)$ is imposed by some experimental measurements or some physical constraints at some densities.
The uncertainties of the correlations will be extracted from the impact of the higher order parameters not included in the correlation itself. The effect of the latter terms will be estimated from a set of $\sim$50 realistic EoS models, that have been successfully compared to a large set of observables in the literature~\cite{Margueron2018a}. 
This chosen set comprises Skyrme, Relativistic Mean Field (RMF), Relativistic Hartree-Fock (RHF), as well as many-body perturbation theory (MBPT) based on 7 chiral N3LO EFT interactions~\cite{Drischler2016} ($\chi$EFT 2016), see Ref.~\cite{Margueron2018a} for the complete list and references.
The correlations can be extracted from the different truncation orders defined in Eqs.~(\ref{eq:esat2})-(\ref{eq:esym4}), as well as from the set of realistic functionals. 
Comparing the results of these different correlations will show the relative importance of the high order parameters in the blurring of the expected correlations.

\section{Results}
\label{sec:results}

In this section, we employ the simple functional~(\ref{eq:energy}) to estimate the strength of various correlations between empirical parameters, such as the well-known correlation between $E_{sym}$ and $L_{sym}$, as well as some other correlations such as the one between $K_{sym}$ and $3E_{sym}-L_{sym}$ recently proposed in Ref.~\cite{Mondal2017}, and we discuss the one between $K_{sat}$ and $Q_{sat}$.

\subsection{Correlation between $E_{sym}$ and $L_{sym}$}
\label{sec:elsym}

\begin{figure}[t]
\begin{center}
\includegraphics[angle=0,width=1.0\linewidth]{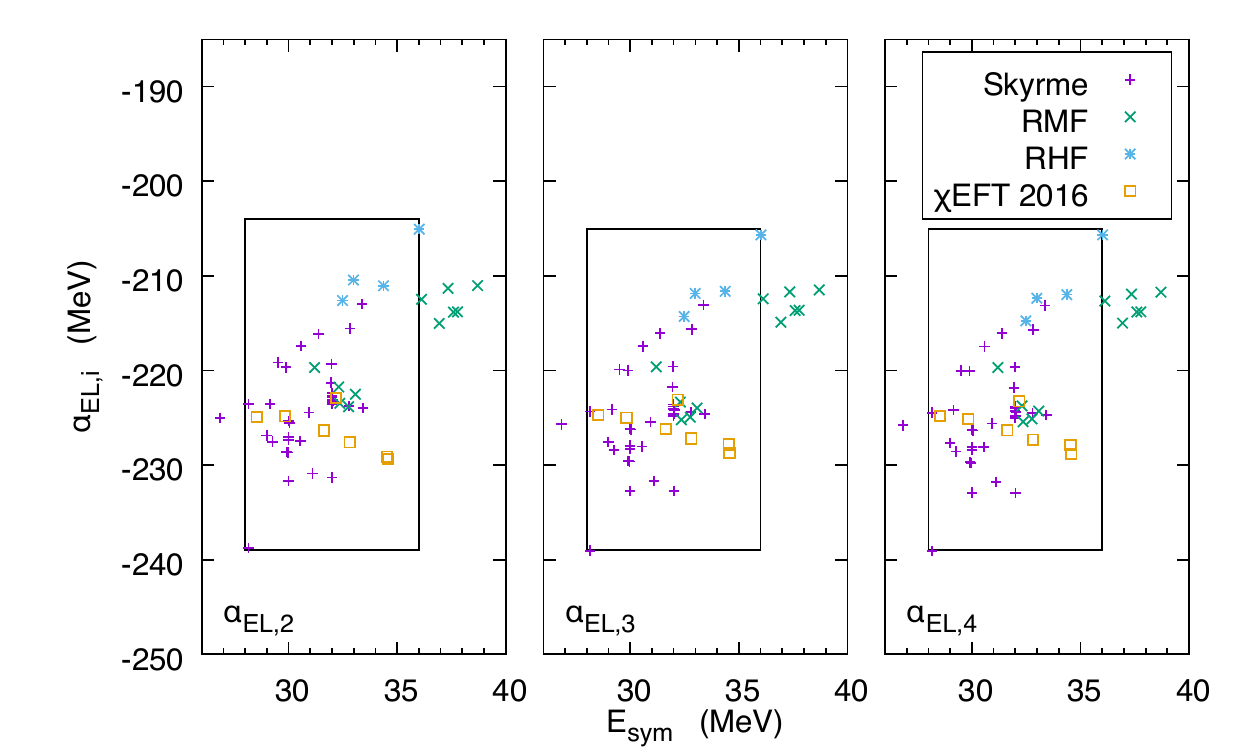}
\end{center}
\caption{(Color online) Expectation values for the parameter $\alpha_{EL,i}$ as function of $E_{sym}$ for a set of different type of nuclear interactions considered in Ref.~\cite{Margueron2018a}: Skyrme, RMF, and RHF. The points labelled $\chi$EFT 2016 stand for the MBPT based on N3LO EFT interaction~\cite{Drischler2016}. See text for more details.
}
\label{fig:EL1} 
\end{figure} 

From the analysis of the giant dipole resonance (GDR) of $^{208}$Pb, a well-constrained estimate of $e_{sym}$ at $n_a=0.1\approx\nicefrac{2}{3}n_{sat}$~fm was proposed~\cite{Trippa2008}.
Original ideas suggesting that finite nuclei data could reveal nuclear properties at the nuclear average density 0.10-0.12~fm$^{-3}$ can also be found in Refs.~\cite{Furnstahl2002,Niksic2008}.
Considering the following condition, $e_{sym}(x=x_a)\equiv E_{sym}^{a}=24.1\pm0.8$~MeV~\cite{Trippa2008,Colo2014}, where $x_a=x(n_a)=-\nicefrac{1}{9}$,
and using Eqs.~(\ref{eq:esym2})-(\ref{eq:esym4}), one can obtain the following correlation between $E_{sym}$ and $L_{sym}$,
\begin{equation}
L_{sym,i} = \beta_{EL} E_{sym} + \alpha_{EL,i}
\label{eq:corrEL}
\end{equation}
with $\beta_{EL}=-x_a^{-1}$ and the value of the $\alpha_{EL,i}$ parameter depends on the truncation order ($i=2$-4) of the Taylor expansion as:
\begin{eqnarray}
\alpha_{EL,2} &=& -x_a^{-1} E_{sym}^{a} - \frac {x_a} {2}K_{sym} \, , \\
\alpha_{EL,3} &=& -x_a^{-1} E_{sym}^{a} - \frac {x_a} {2}K_{sym}-\frac {x_a^2} {6} Q_{sym} \, , \\
\alpha_{EL,4} &=& -x_a^{-1} E_{sym}^{a} - \frac {x_a} {2}K_{sym}-\frac {x_a^2} {6} Q_{sym}-\frac {x_a^3} {24} Z_{sym} \, .
\end{eqnarray}

\begin{figure}[t]
\begin{center}
\includegraphics[angle=0,width=1.0\linewidth]{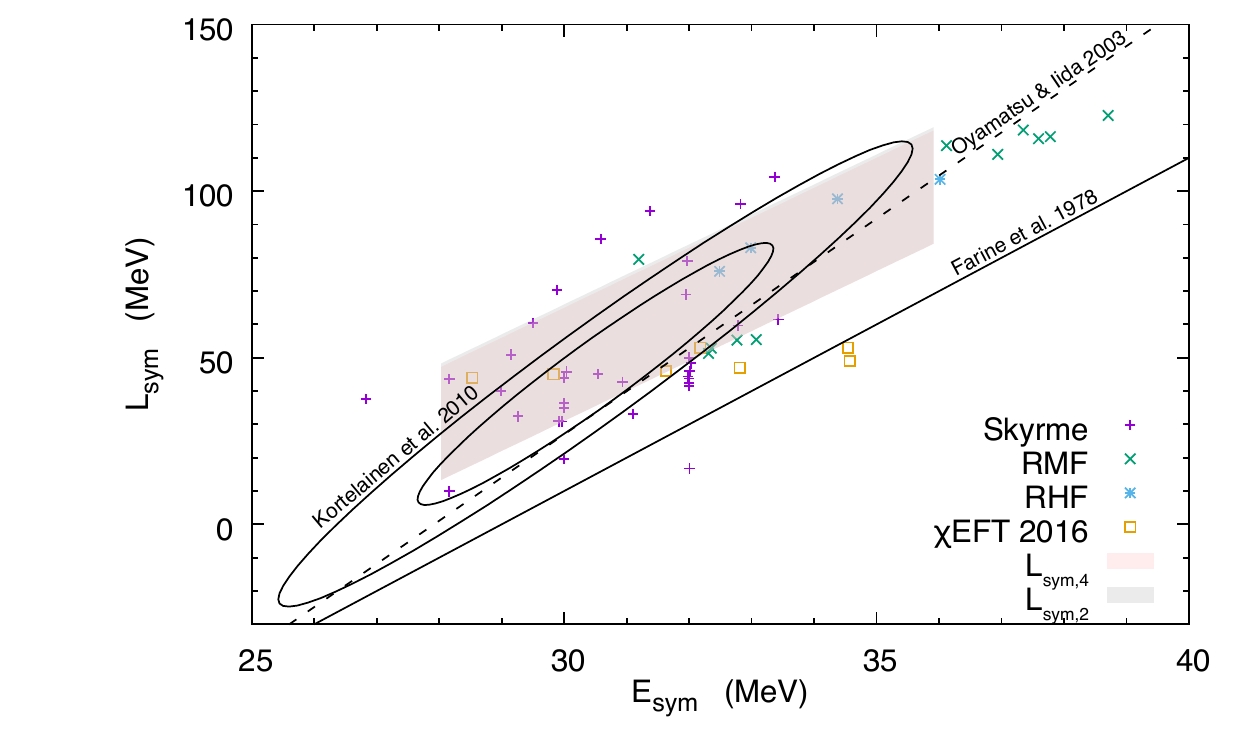}
\end{center}
\caption{(Color online) Correlation between the empirical parameters $L_{sym}$ and $E_{sym}$
for different kind of nuclear interactions described in the text (Skyrme, RMF, RHF, and $\chi$EFT 2016~\cite{Drischler2016}) are also plotted. The bands stands for the uncertainty in the correlation estimated from $\alpha_{EL,i}$ at orders $i=2$-4. The solid line is the correlation obtained by Farine et al. 1978~\cite{Farine1978}, Dashed line by Oyamatsu \& Iida 2003~\cite{Oyamatsu2003}, and the ellipses the 68\% and 95\% confidence intervals of Kortelainen et al. 2010~\cite{Kortelainen2010}.}
\label{fig:EL2} 
\end{figure}

Both $\beta_{EL}$ and $\alpha_{EL,i}$ are affected by some uncertainties. The uncertainty on $\beta_{EL}$ depends on the width of the density domain 
which is effectively explored in the experiment. This information is difficult to evaluate and is not provided from the analysis of the GDR~\cite{Trippa2008}.
In the present case, we therefore fix $\beta_{EL}$ to be $\beta_{EL}=9$ without uncertainty. The uncertainty on the $\alpha_{EL,i}$ parameter explicitly depends on the uncertainty on the high order empirical parameters, which are not strongly constrained by empirical observations. We can estimate this uncertainties by considering a set of $\sim$50 chosen realistic phenomenological functional (Skyrme, RMF and RHF) and many-body perturbation theory (MBPT) based on 7 chiral N3LO EFT interactions~\cite{Drischler2016} ($\chi$EFT 2016), for which the empirical parameters are all given in Ref.~\cite{Margueron2018a}.
Using the predicted empirical parameters for these $\sim$50 functionals, the coefficients $\alpha_{EL,i}$ for the different orders $i=2$-4 are represented in Fig.~\ref{fig:EL1}, as a function of $E_{sym}$. 
If we now select the models for which $28<E_{sym}<36$~MeV~\cite{BALi2013} (rectangles in Fig.~\ref{fig:EL1}), we obtain the following estimation for the coefficients $\alpha_{EL,i}$:
$\alpha_{EL,2}=-221.5\pm 17.5$~MeV, $\alpha_{EL,3}=-222\pm 17$~MeV, and $\alpha_{EL,4}=-222\pm 17$~MeV.
The contribution of the uncertainty in the estimated value of $E_{sym}^{a}$~\cite{Trippa2008} accounts for $\sim$7~MeV of the total uncertainty in $\alpha_{EL,i}$, while the rest of the uncertainty accounts for the contribution of the high order parameters estimated from the selected models. The estimation of $\alpha_{EL,i}$ at different orders $i$ closely agree, indicating that the value of $\alpha_{EL,i}$ is essentially determined by $E_{sym}^{a}$, $n_c$ and the isovector incompressibility $K_{sym}$, while the higher order parameters play a negligible role.
A better knowledge of the empirical parameter $K_{sym}$ will therefore lead to an improvement of the $E_{sym}$-$L_{sym}$ correlation.

The correlation~(\ref{eq:corrEL}) between $E_{sym}$ and $L_{sym}$ is shown in Fig.~\ref{fig:EL2} varying the coefficient $\alpha_{EL,i}$ within the boundaries obtained from the analysis of Fig.~\ref{fig:EL1}.
The gray band corresponds to $\alpha_{EL,2}$ and the pink one to $\alpha_{EL,4}$.  
For comparison, the $E_{sym}-L_{sym}$ correlation for the $\sim$50 considered models is also plotted in Fig.~\ref{fig:EL2}. 
There is a good overlap between our predicted correlation band and the values $E_{sym}$-$L_{sym}$ predicted by the $\sim$50 considered models, indicating that i) our simple analytical model for the symmetry energy~(\ref{eq:esym2})-(\ref{eq:esym4})  can efficiently map the $E_{sym}$-$L_{sym}$ correlation and ii) our error estimate for the $E_{sym}$-$L_{sym}$ correlation is satisfactory.
Considering only the $\sim$50 models sampling, the correlation coefficient between $L_{sym}$ and $E_{sym}$ is found to be $\sim$0.80 and $\sim$0.55 if we consider the reduced sample of models for which  $28<E_{sym}<36$~MeV.
In summary, the dispersion of the $E_{sym}$-$L_{sym}$ correlation can be understood as partially coming from the experimental uncertainty in $E_{sym}^{a}$ and partially due to the uncertainty on the poorly known empirical parameter $K_{sym}$.

The $E_{sym}$-$L_{sym}$ correlation was discussed in earlier works, so we report in Fig.~\ref{fig:EL2} on other $E_{sym}$-$L_{sym}$ correlations, as in Ref.~\cite{Lattimer2013}. One of the first studies of the correlation between $E_{sym}$ and $L_{sym}$ from a Skyrme mass formula was presented in Ref.~\cite{Farine1978}, called 'Farine et al. 1978 in Fig.~\ref{fig:EL2}. Based on a macroscopic nuclear model, the $E_{sym}$-$L_{sym}$ correlation was later re-examined~\cite{Oyamatsu2003}, 'Oyamastu \& Iida 2003'.
The 68\% and 95\% confidence intervals of 'Kortelainen et al. 2010'~\cite{Kortelainen2010} are also plotted in Fig.~\ref{fig:EL2}.
Our analysis agrees well with more recent investigations: the correlation coefficient was found to be $\sim$0.71 in Ref.~\cite{Kortelainen2012} and 0.9-0.95 in Ref.~\cite{Nazarewicz2014} (the variation in the correlation coefficient reflects the dispersion of the models). Based on a different sampling of models a sizeable $E_{sym}$-$L_{sym}$ correlation was also found in Ref.~\cite{Ducoin2011}. Concerning the experimental probe to be chosen to determine $E_{sym}^{a}$, an alternative choice was proposed from an analysis of the isobaric analog state (IAS) and neutron skin radius~\cite{Danielewicz2014}.

In summary, we found a satisfactory agreement between the $E_{sym}$-$L_{sym}$ correlation suggested from our analysis and the dispersion of the $\sim$50 models considered here, as well as with previous investigations. In addition, our analysis suggests that the better knowledge of the empirical parameter $K_{sym}$ will reduce the blurring of the correlation.

\subsection{Correlation between $K_{sym}$ and $3E_{sym}-L_{sym}$}

A recent analysis of 500 different density functional models has revealed a general correlation between the empirical parameter $K_{sym}$ and the linear combinaison $3E_{sym}-L_{sym}$ as,
\begin{equation}
K_{sym} = \beta (3E_{sym}-L_{sym}) + \alpha \, ,
\label{eq:corrK}
\end{equation}
where the fit gives $\alpha=66.80\pm 2.14$~MeV and $\beta=-4.97\pm0.07$~MeV~\cite{Mondal2017}. 
The origin of such a correlation was however not explained in Ref.~\cite{Mondal2017}.
In this section, we propose a simple explanation for the correlation~(\ref{eq:corrK}).

Defining the energy of neutron matter (NM) from Eq.~(\ref{eq:energy}) as $e_{NM}(n)=e_{sat}(n)+e_{sym}(n)$, we impose the very general constraint that the neutron energy per particle should be zero at zero density,
\begin{eqnarray}
e_{NM}(x=-\nicefrac{1}{3})&=&0 \hbox{ MeV} \, ,
\end{eqnarray}
which gives the following linear combination among the empirical parameters,
\begin{eqnarray}
e_{NM,4}(x=-\nicefrac{1}{3})&=&0=E_{sat}+E_{sym}-\frac 1 3 L_{sym} + \nonumber \\
&& \hspace{-3cm}\frac 1 {18}(K_{sat}+K_{sym} ) - \frac 1 {162} Q_{NM} + \frac 1 {1944} Z_{NM} + ... \, ,
\end{eqnarray}
where $Q_{NM}=Q_{sat}+Q_{sym}$ and $Z_{NM}=Z_{sat}+Z_{sym}$.
This condition can be expressed as a correlation between $K_{sym}$ and the linear combinaison $3E_{sym}-L_{sym}$ -- which naturally appears here -- as,
\begin{equation}
K_{sym,i} = \beta_{K_{sym}} (3E_{sym}-L_{sym}) + \alpha_{K_{sym},i} \, .
\label{eq:ksymi}
\end{equation}
with $\beta_{K_{sym}}=-6$ and $\alpha_{K_{sym},i}$ at different orders, 
\begin{eqnarray}
\alpha_{K_{sym},2} &=& -18E_{sat}-K_{sat} \, , \\
\alpha_{K_{sym},3} &=& -18E_{sat}-K_{sat}+Q_{NM}/9 \, , \\
\alpha_{K_{sym},4} &=& -18E_{sat}-K_{sat}+Q_{NM}/9-Z_{NM}/108 \, .
\end{eqnarray}
$\alpha_{K_{sym},2}$ is the  expectation for the constant $\alpha_{K_{sym},i}$ assuming a Taylor expansion up to second order only, while $\alpha_{K_{sym},3}$ and $\alpha_{K_{sym},4}$
take into account the uncertainties induced by the unknown higher orders terms.

\begin{figure}[t]
\begin{center}
\includegraphics[angle=0,width=1.0\linewidth]{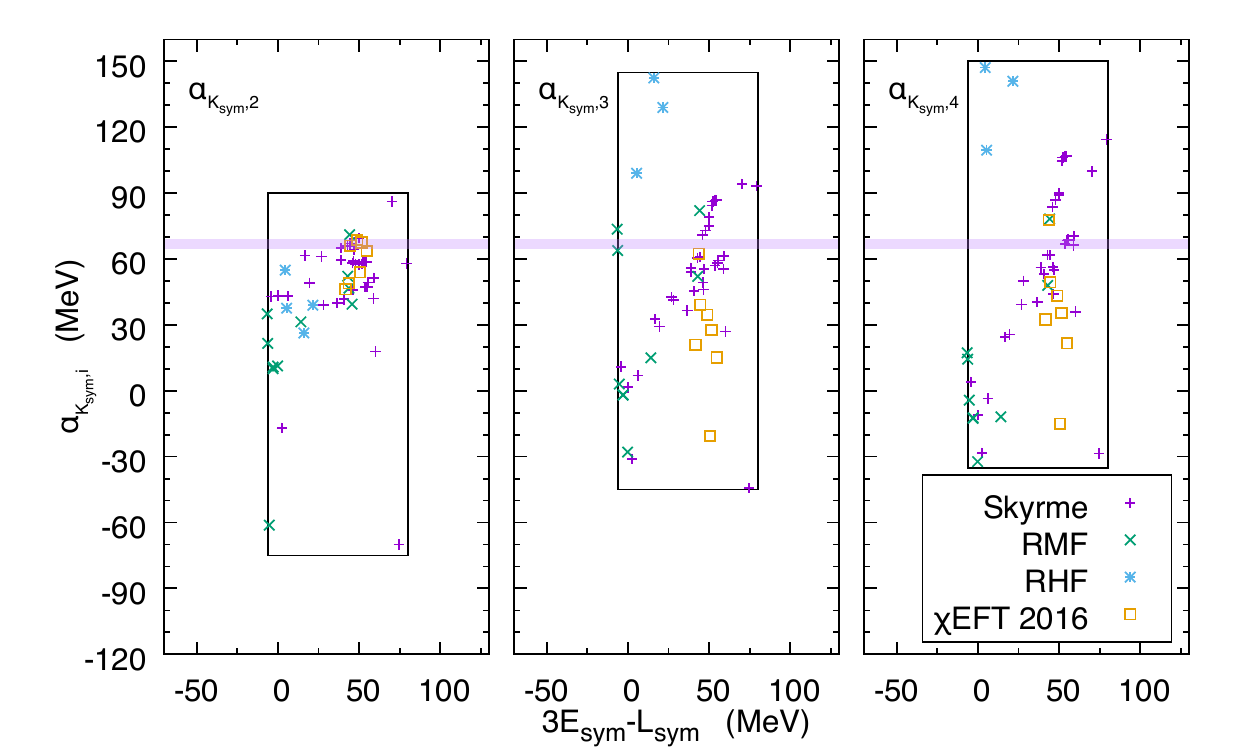}
\end{center}
\caption{(Color online) Expectations values for $\alpha_{K_{sym},i}$ as function of the linear combinaison $3E_{sym}-L_{sym}$ for a set of different types of energy functionals, as in Fig.\ref{fig:EL1}.
The value for $\alpha$ given in Ref.~\cite{Mondal2017} is also shown (purple band).}
\label{fig:alpha} 
\end{figure}

Assuming $E_{sat}=-16\pm0.5$~MeV and $K_{sat}=230\pm20$~MeV~\cite{Khan2012,Bertsch2017,Margueron2018a}, one can get a rough estimation for $\alpha_{K_{sym},2} \approx 58\pm30$~MeV, which is compatible with the coefficient $\alpha$ fitted in Ref.~\cite{Mondal2017}.
As in the previous section, the average values of  $\alpha_{K_{sym},i}$ ($i=2$-4) and their uncertainties can be estimated from a set of the same $\sim$50 functionals. The result is given  in Fig.~\ref{fig:alpha} as function of the linear combinaison $3E_{sym}-L_{sym}$.

Note that, at variance with the $E_{sym}$-$L_{sym}$ correlation, there is no "experimental" uncertainty on the energy or the density here. The blurring of the correlation  can only come from the role of the high order empirical parameters. Comparing the results for different truncation orders in Fig.~\ref{fig:alpha} we can see that there is a noticeable correction in $\alpha_{K_{sym},i}$ induced by the third order parameter $Q_{NM}$, while the correction induced by the forth order parameter is less important.
Rectangular boxes are also drawn into Fig.~\ref{fig:alpha}. 
They comprise the results obtained from models which correspond to values for the linear combinaison $-6\leq 3E_{sym}-L_{sym}\leq 80$. 
This interval is obtained considering the conservative estimation for $E_{sym}$ and $L_{sym}$: $28\leq E_{sym}\leq 36$ and $30\leq L_{sym} \leq 90$~\cite{BALi2013,Margueron2018a}.
Note that this box contains all our $\sim$50 models, but it is possible to find other models out of this box, see for instance Ref.~\cite{Mondal2017}.
We deduce the following uncertainties for $\alpha_{K_{sym,}i}$ at different orders:
$\alpha_{K_{sym},2}=8\pm83$~MeV, $\alpha_{K_{sym},3}=50\pm95$~MeV, and $\alpha_{K_{sym},4}=58\pm93$~MeV.

\begin{figure}[t]
\begin{center}
\includegraphics[angle=0,width=1.0\linewidth]{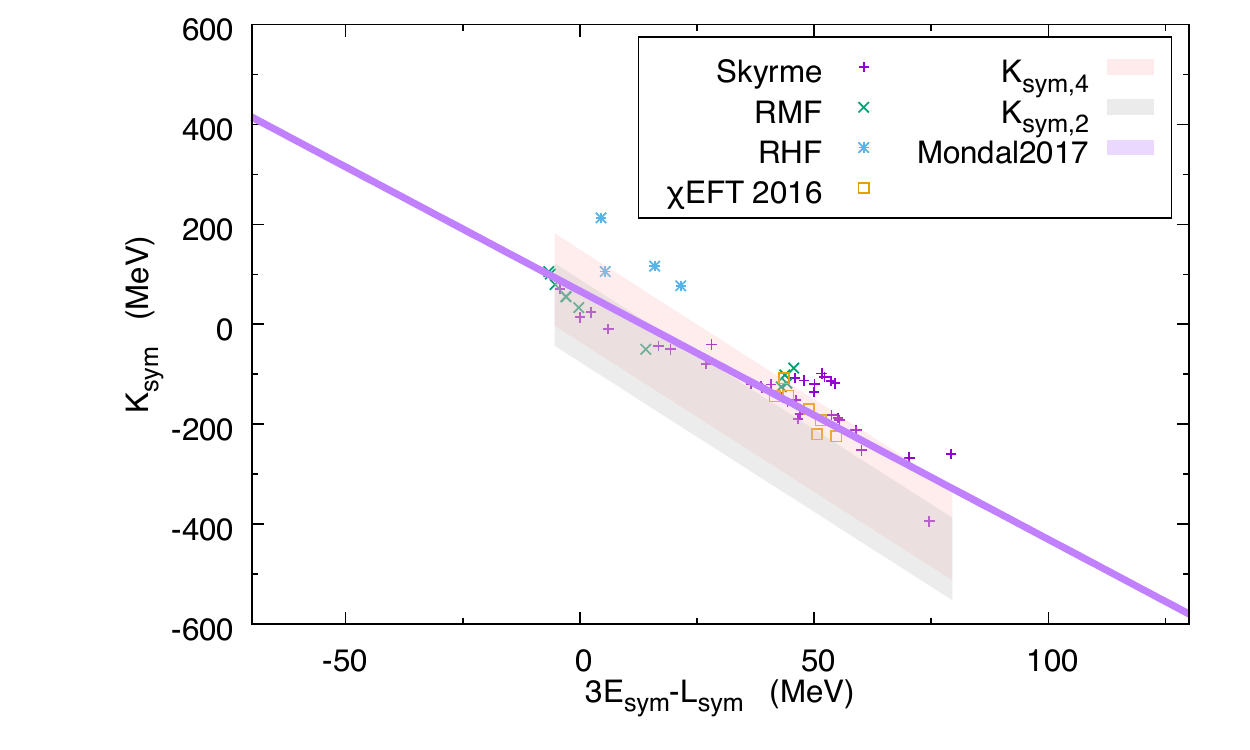}
\end{center}
\caption{(Color online) Correlation between the empirical parameters $K_{sym}$ and the variable $3E_{sym}-L_{sym}$
for different kind of nuclear interactions, as in Fig.~\ref{fig:EL2}.
The fit from Ref.~\cite{Mondal2017} is shown as well as our analytical expression taken with different order corrections.}
\label{fig:Ksym} 
\end{figure}

We now represent in Fig.~\ref{fig:Ksym} the correlation between $K_{sym}$ and $3E_{sym}-L_{sym}$ within different cases: the different bands correspond to the correlation~(\ref{eq:ksymi}) at different orders $i=2$, 4 while the thinner band labelled 'Mondal2017' shows the result of the fit from Ref.~\cite{Mondal2017}. The points show the position of the $\sim$50 models as in previous figures.

We remark from Fig.~\ref{fig:Ksym} that the correlation~(\ref{eq:ksymi}) deduced from the condition $e_{NM}(x=-\nicefrac{1}{3})=0$~MeV is  very consistent with the behavior of the $\sim$50 models as well as with the fit from Ref.~\cite{Mondal2017} (Mondal2017) where a larger number of models has been considered.
The dispersion in the fit Mondal2017 is however smaller than in our case, and Fig.~\ref{fig:Ksym} shows that many models are indeed out of the fit Mondal2017.
It was already clear from the results presented in Ref.~\cite{Mondal2017} that the dispersion of the fit was underestimating the one of the model sample.
The estimation of the dispersion of the correlation~(\ref{eq:ksymi}) obtained in our case is closer to the one of our models, as shown in Fig.~\ref{fig:Ksym}.
It seems to reproduce also very well the larger sample of model shown in Ref.~\cite{Mondal2017}.
The correlation~(\ref{eq:ksymi}) and its dispersion at orders $i=3$-4 are very close -- we have therefore represented only $i=4$ -- but they are slightly different from the correlation at order $i=2$, where most of the dispersion is generated by the uncertainty in $K_{sat}$. 
The impact of adding the skewness parameter $Q_{NM}=Q_{sat}+Q_{sym}$ (at order $i=3$) is to shift up the correlation, improving the overlap with the $\sim$50 models.
We can therefore conclude that while most of the correlation~(\ref{eq:ksymi}) relies on the knowledge of $E_{sat}$ and $K_{sat}$ (isoscalar parameters), the role of the skewness parameter $Q_{NM}$ is also important to better reproduce the datum while the higher order parameter $Z_{NM}$  can here be neglected.

Given the rather large dispersion in the correlation~(\ref{eq:ksymi}), the correlation coefficient extracted directly from the points largely depends on the model sampling.
While it is found to be $-0.91$ for the $\sim$50 models considered here, it decreases to $-0.48$ if we reduce the range of the x-axis to $30$-$60$~MeV as suggested by $\chi$EFT analyses~\cite{Tews2017,Drischler2016}: $28\leq E_{sym}\leq 36$~MeV and $40\leq L_{sym} \leq 60$~MeV.

To summarize, the correlation proposed in Ref.~\cite{Mondal2017} can be related to the very general condition $e_{NM}(x=-\nicefrac{1}{3})=0$~MeV and the dispersion is related to our uncertainty in the empirical parameter $K_{sat}$ in symmetric matter as well the skewness parameter $Q_{NM}$ in neutron matter, which is almost unknown. In the present case, we guesstimated its value and dispersion from a set of "realistic" models. The correlation coefficient depends largely on the dispersion of the model prediction for the variable $3E_{sym}-L_{sym}$.
The correlation proposed in Ref.~\cite{Mondal2017} is therefore very interesting but not very constraining within the present knowledge of nuclear physics.

\subsection{Correlation between $K_{sat}$ and $Q_{sat}$}

We have discussed in the previous sections that the physical correlations between empirical EoS parameters, such as the well-known correlation between 
$E_{sym}$ and $L_{sym}$, or the more recently observed \cite{Mondal2017} correlation between  $K_{sym}$ and $3E_{sym}-L_{sym}$, are largely blurred by our present poor knowledge on the high order parameters, both in the isoscalar and in the isovector sector. 

A way to reduce this uncertainty could be to pin down these high order parameters from some existing correlation with the low order ones, which are more effectively constrained by experimental data.  
For this reason we examine in the present section the correlation between $K_{sat}$ and $Q_{sat}$.
From the observation of our representative set of $\sim$50 EoS models, it comes out that this correlation is weak. 
We found for the  considered models a coefficient of 0.52, and if we reduce the sampling to the more realistic models for which $210<K_{sat}<250$~MeV, then the correlation coefficient drops down to 0.23.
We want here to understand what are the physical reasons of such an absence of correlation.

The density dependent  incompressibility in symmetric matter (SM) is defined as $K_v(n)=9\,n\,\partial^2\epsilon/\partial n^2\,(\delta=0)$, with $\epsilon(n)=n \,e_{sat}(n)$.  
Using the Taylor expansion Eqs.(\ref{eq:esat2})-(\ref{eq:esat4}), it reads,
\begin{eqnarray}
  \frac{K_{v}(x)}{1+3x} &=& \left ( {1+9x}\right ) K_{sat} +x  \left ({1+6x}\right )  Q_{sat}
+\frac{x^2}{2}  \left ( {1+5x}\right ) Z_{sat} + ... \, \nonumber \\
\label{eq:Kv}
\end{eqnarray}

It was recently observed that the incompressibility $K_v$  calculated for different models crosses at a density of about $n_c = (0.71\pm0.01) n_{sat} = 0.114\pm0.002$~fm$^{-3}$, for a value which is 
$K_{v,c}=37\pm 8$~MeV~\cite{Khan2012}, where the systematic dispersion between Skyrme and Gogny type models is included in the error-bars.
The reason of this behavior was understood from the fact that these different models have been calibrated to reproduce the experimental value of the GMR, which provides a constraint at the average density of finite nuclei. It is therefore not surprising that the value of the crossing density $n_c$ is close to the average density in nuclei $n_a$ previously introduced in Sec.~\ref{sec:elsym}. 
Indeed the experimental value of the GMR turns out to be well correlated with the parameter $M_c$ defined as
$M_c=3 n_c \, \partial K_{v}/ \partial n\,(n=n_c)$, and an experimental value for $M_c=1050\pm100$~MeV was deduced from the correlation of this parameter with the ISGMR energy of Sn and Pb~\cite{Khan2012}.

The parameter $M_c$ can be deduced from Eq.~(\ref{eq:Kv}) as,
\begin{eqnarray}
 \frac{M_{c}}{1+3x_c} &=&  6\left( 2 + 9x_c \right ) K_{sat} + \left ({1+18x_c+54x_c^2}\right ) Q_{sat}\nonumber \\
&&\hspace{1cm}+  x_c \left ({1+12x_c+30x_c^2}\right ) Z_{sat} + ... \, ,
\label{eq:Mc}
\end{eqnarray}
which, for a typical value $x_c\sim-0.1$, gives
\begin{eqnarray}
M_{c} \approx  4.6 \,K_{sat} - 0.18 \, Q_{sat} - 0.007 \, Z_{sat} + ... \, .
\label{eq:Mcapprox}
\end{eqnarray}
There is therefore still a very strong correlation between $M_c$ and $K_{sat}$, and the influence of $Q_{sat}$ is non-negligible but it remains small ($Q_{sat}$ is not well known, but typical values of nuclear models are of the order of a few $\pm K_{sat}$~\cite{Margueron2018a}).
For densities below $n_{sat}$, the coefficient in front of $Q_{sat}$ is bounded between $-0.5$ and $1$ and is even passing by zero at two densities: $n\approx0.12$-$0.13$~fm$^{-3}$ and $n\approx0.03$-$0.04$~fm$^{-3}$.
The impact of $Q_{sat}$ on the parameter $M_c$ is therefore quenched around $n_c$, explaining why the coefficient in front of $Q_{sat}$ in Eq.~(\ref{eq:Mcapprox}) is so small.
Since the energy of the ISGMR is very well correlated with $M_c$~\cite{Khan2012,Khan2013}, we can understand \textsl{a posteriori} why there is still a good correlation between the energy of the ISGMR and the empirical parameter $K_{sat}$. Such a correlation has been widely used to estimate the value of $K_{sat}$ from experimental measurement of the ISGMR since the seminal work of Blaizot~\cite{Blaizot1980}, see for instance Refs.~\cite{Colo2004,Colo2014,Khan2012}.
The impact of the other empirical parameters $Q_{sat}$ and $Z_{sat}$ remains small, but they are important for accurate and model independent determination of empirical parameters~\cite{Khan2012,Khan2013}.
As a consequence, a better knowledge of $Q_{sat}$ is necessary to reduce the uncertainty in the determination of $K_{sat}$.

Even if the impact of $Q_{sat}$ is small in Eq.~(\ref{eq:Mc}), it is still possible to use Eq.~(\ref{eq:Mc}) to express the following correlation between $K_{sat}$ and $Q_{sat}$,
\begin{equation}
Q_{sat,i} = \beta_{KQ} K_{sat} + \alpha_{KQ,i} \,
\label{eq:corrKQ2}
\end{equation}
where
\begin{equation}
\beta_{KQ} = -\frac{6(2 + 9x_c) }{ 1  + 18 x_c+54x_c^2} \, ,
\end{equation}
and
\begin{eqnarray}
\alpha_{KQ,3} = \alpha_1 M_c \, , \hspace{0.5 cm} \alpha_{KQ,4} &=& \alpha_1 M_c+\alpha_2 Z_{sat}  \, ,
\end{eqnarray}
with
\begin{eqnarray}
\alpha_1 &=& \frac{1}{\left (1+3x_c\right )\left ( 1  + 18 x_c+54x_c^2\right )   } \, , \\
\alpha_2 &=& -  x_c \frac{1+12x_c+30x_c^2}{1+18x_c+54x_c^2} \, . 
\end{eqnarray}
Considering the uncertainty in $n_c$, we obtain $\beta_{KQ}=29\pm4$, $\alpha_1=-6.06\pm0.57$ and $\alpha_2=-0.0505\pm0.012$.
Considering in addition the uncertainty in $M_c$, we find $\alpha_{KQ,3}=-6300\pm1200$~MeV.

\begin{figure}[t]
\begin{center}
\includegraphics[angle=0,width=1.0\linewidth]{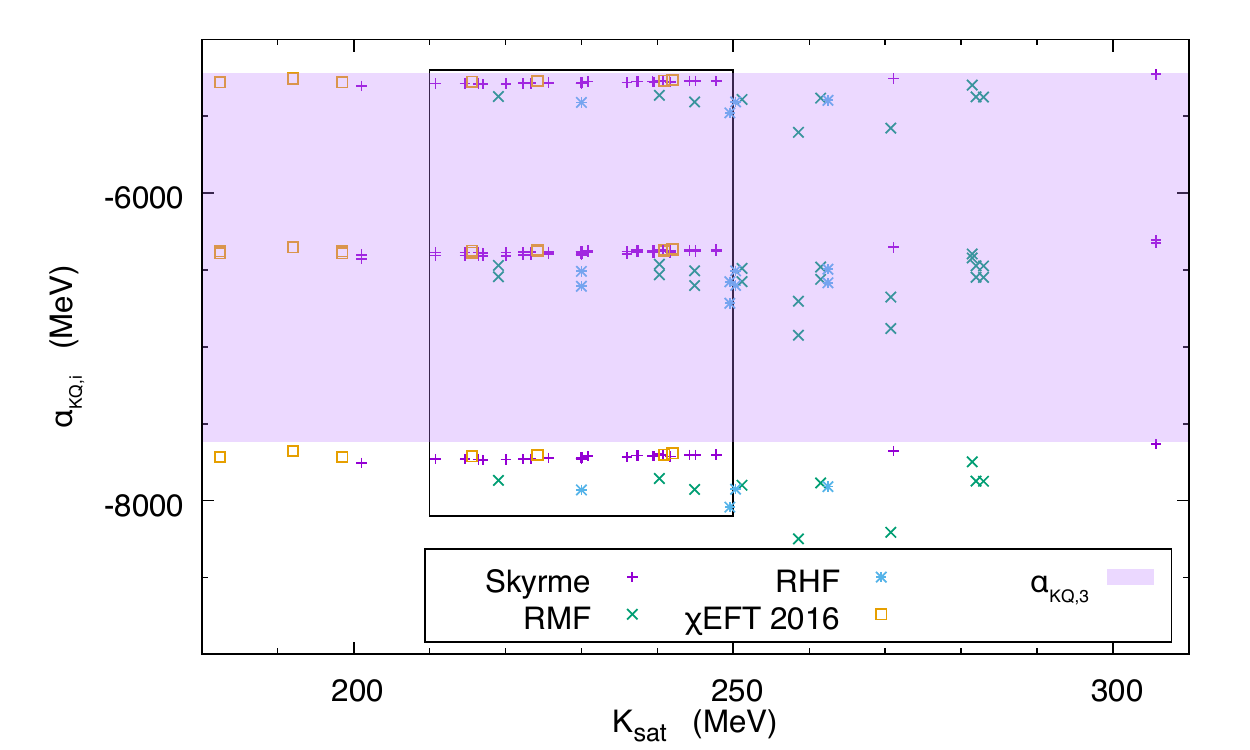}
\end{center}
\caption{(Color online) Expectations values for $\alpha_{KQ,i}$ as function of $K_{sat}$ for a set of different type of nuclear interactions, as in Fig.\ref{fig:EL1}. }
\label{fig:KQ3} 
\end{figure}

The value of  $\alpha_{KQ,4}$ is a more difficult to calculate since it  implies the parameter $Z_{sat}$ which is unknown.
Similar to the strategy of the previous sections, we evaluate $\alpha_{KQ,4}$ from our set of $\sim$50 nuclear models.
The result is shown in Fig.~\ref{fig:KQ3}, and a rectangle sets the most realistic boundaries under the assumption that $210<K_{sat}<250$~MeV,
giving $\alpha_{KQ,4}=-6650\pm1450$~MeV. The values allowed for $\alpha_{KQ,3}$ are also shown in Fig.~\ref{fig:KQ3}.

\begin{figure}[t]
\begin{center}
\includegraphics[angle=0,width=1.0\linewidth]{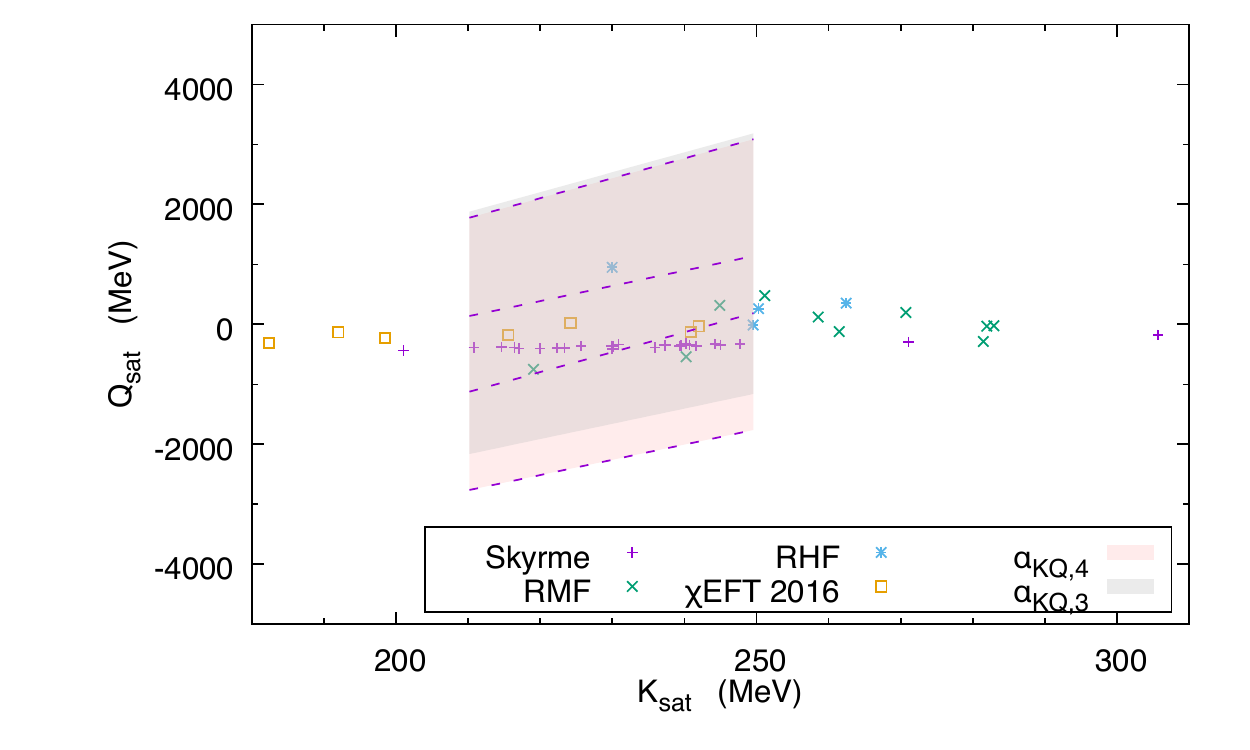}
\end{center}
\caption{(Color online) Correlation between the empirical parameters $Q_{sat}$ and $K_{sat}$
for different kind of nuclear interactions, as in Fig.\ref{fig:EL2}.}
\label{fig:KQ4} 
\end{figure}

Combining the uncertainties in $\beta_{KQ}$, $\alpha_{KQ,3}$ and $\alpha_{KQ,4}$, we compare the correlation (\ref{eq:corrKQ2}) to the values of the $\sim$50 nuclear models in Fig.~\ref{fig:KQ4}.
We can see that the estimated band is wide enough to contain the predictions of all $\sim$50 nuclear models.
At variance with the previous correlations, the band of the $Q_{sat}$-$K_{sat}$ correlation is even wider than the actual spreading among the models.
From our analysis, this wide width comes from both the uncertainty in the crossing density $n_c$ and in the parameter $M_c$.
Since $\alpha_1\approx 6$, the uncertainty in $M_c$ is largely amplified for $Q_{sat}$.

There could be at least two reasons why the $\sim$50 models seem to have a smaller width than our prediction based on Eq.~(\ref{eq:Mc}).
The first reason is that there may be another constraint satisfied by the $\sim$50 models which tights the band's width, and which is not included in our analysis.
One may think for instance of the surface energy which provides constraint in the density dependence of the energy per particle and which is not included in our analysis.
It is however also known that phenomenological models exhibit spurious correlations in the $Q_{sat}$-$K_{sat}$ diagram, see for instance Refs.~\cite{Khan2012,Margueron2018a}.
So the other reason could be that the dispersion among the $\sim$50 models is artificially smaller than it should be in reality.

\begin{figure*}[tb]
\begin{center}
\includegraphics[angle=0,width=0.8\linewidth]{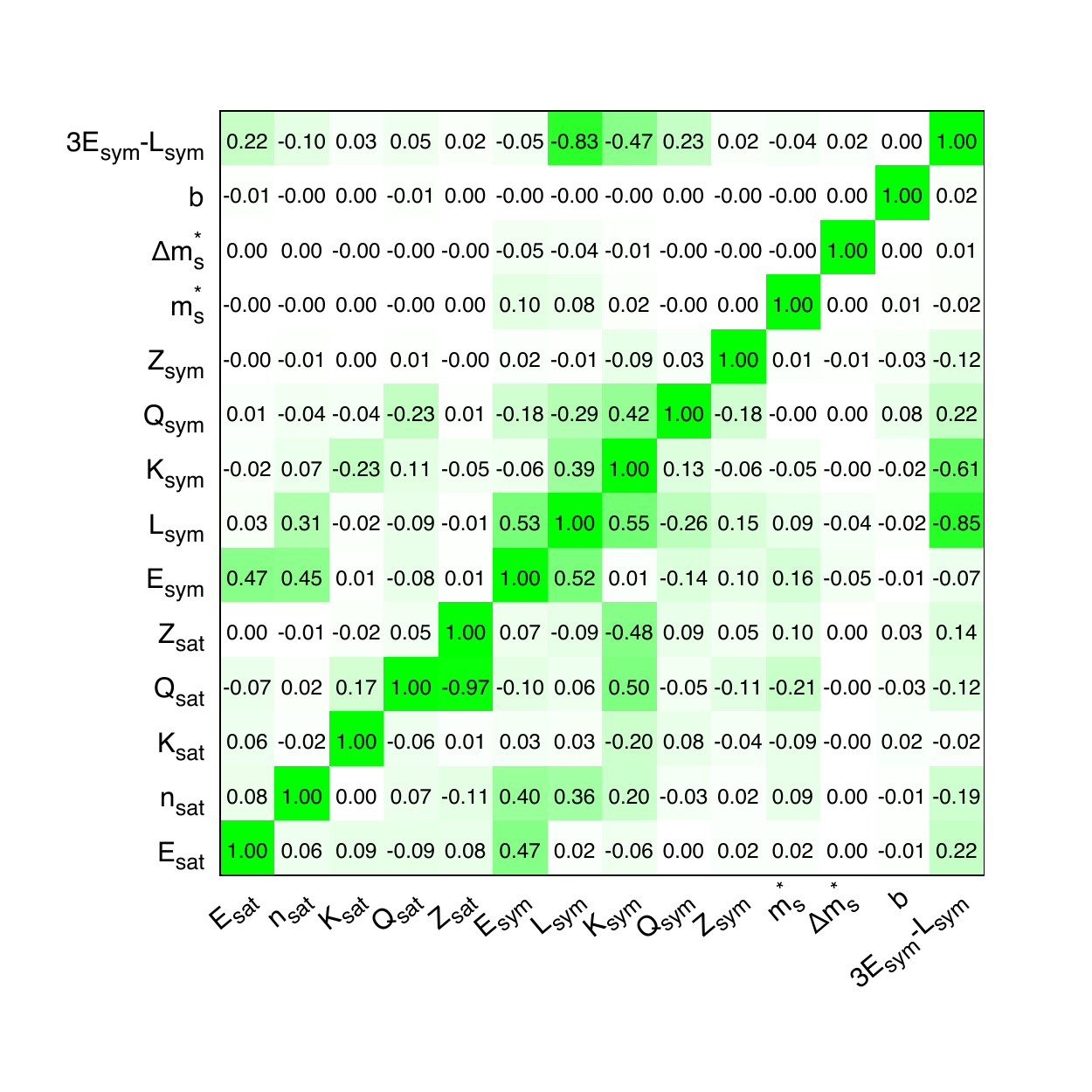}
\end{center}
\vspace{-1.5cm}
\caption{(Color online) Correlation among the empirical parameters, completed with the effective mass $m_{sat}^*$, the effective mass splitting $\Delta m_{sat}^*$, the parameter $b$ and the variable $3E_{sym}-L_{sym}$. The correlation above the diagonal corresponds to the fit of the $\chi$EFT~\cite{Drischler2016} predictions only, while below the diagonal, the stability and causality conditions are added up to $0.4$~fm$^{-3}$. See text for discussion.}
\label{fig:MM} 
\end{figure*}

Let us mention that we have also explored the $Q_{sat}$-$K_{sat}$ correlation generated by the crossing value $K_{v,c}$, as well as the one emerging from the spinodal condition -- though there is no experimental measurement of it. Our conclusion on the band width in these two cases are the same as the one found here based on the experimental measurement of $M_c$.

In summary, while the width of the $Q_{sat}$-$K_{sat}$ correlation is quite large from our analysis, we are able to determine an upper and lower bound in this correlation simply related to the experimental determination of the parameter $M_c$. The reason for the $Q_{sat}$-$K_{sat}$ correlation to be very  weak is the small contribution of the parameter $Q_{sat}$ to the incompressibility in the region of densities around $n_c$. In the future, it would be interesting to estimate the reduction of the width of the  $Q_{sat}$-$K_{sat}$ correlation induced by additional constraints, such as for instance the one provided by the surface energy.

\section{Meta-modelling analysis of the correlations}
\label{sec:MM}

In Sec.~\ref{sec:results}, we have employed the simple and analytical model presented in Sec.~\ref{sec:def} to perform various correlation analyses among empirical parameters, considering a given experimental constraints: $e_{sym}(n=\nicefrac{2}{3}n_{sat})$ for the $E_{sym}$-$L_{sym}$ correlation, $e_{NM}(n=0)$ for the $K_{sym}$-$(3E_{sym}-L_{sym})$ correlation and $M_c$ for the $K_{sat}$-$Q_{sat}$ correlation. These correlations are blurred by the presence of other empirical parameters, which aren't known for most of them. The estimation of the importance of the blurring is therefore not an easy task. In order to circumvent this issue, we used a set of $\sim$50 nuclear functionals to estimate the dispersion among the unknown empirical parameters. The parameters of these $\sim$50 functionals have been optimized on different nuclear structure data, meaning that we can consider that experimental constraints from low energy nuclear experiments are somehow implicitly accounted in the choice of parameters. These low density constraints are however not always sufficient to pin down the behavior of high order parameters, both via direct measurement or via the exploitation of correlations with low order parameters. Indeed, the correlations involving high order parameters observed in existing phenomenological nuclear models are essentially induced by the assumed functional form and do not reflect physical constraints. For this reason, the correlations analyzed in Sec.~\ref{sec:results} may still potentially contain some model dependence.

To overcome this problem, we have recently proposed~\cite{Margueron2018a,Margueron2018b} a meta-modelling formulation of the EoS employing i) a functional form flexible enough to be able to reproduce within its parameter space, most relativistic and non-relativistic functionals, including ab-initio ones; ii) no \textsl{a priori} correlation among the empirical parameters -- such that we can consider a portion of the parameter space which is not explored by existing models; iii) an \textsl{a posteriori} filtering of the huge parameter space with basic physical requirements (stability and causality) and the existing constraints from \textsl{ab initio} approaches, such as the MBPT based on $\chi$EFT interactions~\cite{Drischler2016}.

We consider in this section the metamodel -- version ELF-c -- of Ref.~\cite{Margueron2018a} which is determined from a given set of empirical parameters, see Eqs.(\ref{eq:empIS})-(\ref{eq:empIV}), from the effective mass $m_{sat}^*$ defined at $n_{sat}$ in symmetric matter, from the effective mass splitting $\Delta m_{sat}^*/m = m_{n}^*/m-m_{p}^*/m$ defined at $n_{sat}$ in neutron matter, and from the parameter $b$ which incorporates, at low density, the effects of the neglected high order terms in the series expansion, see Ref.~\cite{Margueron2018a} for more details. These parameters are sampled as in Ref.~\cite{Margueron2018b} and they are first filtered against the MBPT predictions based on $\chi$EFT interactions~\cite{Drischler2016} in symmetric and neutron matter, in a similar way as it has been done in Ref.~\cite{Tews2018}.
Since we may want to control the behaviour of the selected models above saturation density -- within a reasonable range -- we have additionally imposed the stability and causality condition up to 0.4~fm$^{-3}$.

We have calculated the correlation coefficient among the 13 parameters of the model, plus the combinaison $3E_{sym}-L_{sym}$, for the following two selection conditions: i) the models selected only from the MBPT predictions in symmetric and neutron matter based on six chiral EFT interactions~\cite{Drischler2016}, and ii) the models additionally filtered against stability and causality. The bayesian selection mentioned in i) assumes that the theoretical MBPT predictions could be used in the definition of a likelihood probability where the theoretical centroid and uncertainty for the binding energy and the baryon pressure define a $\chi^2$. Each model set is weighted with the likelihood probability $p=\exp[-\chi^2 / (2 N_{dof})]$ where $N_{dof}=N_{tot}-13$, $N_{tot}=32$ for 8 density points from 0.04 to 0.20~fm$^{-3}$. Note that more evolved bayesian analyses could be perform, see for instance Ref.~\cite{Melendez:2017}. While the details of the marginalized posterior probabilities certainly depend on the bayesian prescription, the gross correlation properties shown in this study are much less impacted.

The results are shown in Fig.~\ref{fig:MM}, where the correlation coefficients above the diagonal are obtained from the selection condition i), and the ones below the diagonal from the selection condition ii).
There is a general agreement for the correlation coefficients obtained from conditions i) and ii), with some exceptions. For instance, the $Q_{sat}$-$Z_{sat}$ correlation is very weak in the case i) while it is very large in the case ii). It is simply due to the stability and causality conditions which bring strong constraints above saturation density, as expected. $K_{sym}$ is more correlated with $Q_{sat}$ and $Z_{sat}$ in the case ii) than in the case i). The $K_{sym}$-$Q_{sym}$ correlation is weaker in case ii) compared to case i). Despite these few exceptions, the correlation coefficients are rather stable and independent of the additional filtering against stability and causality.

The $E_{sym}$-$L_{sym}$, the $K_{sym}$-$(3E_{sym}-L_{sym})$, and the $K_{sat}$-$Q_{sat}$ correlation coefficients shown in Fig.~\ref{fig:MM} essentially confirm our previous analysis in Sec.~\ref{sec:results}. The correlation coefficient for the $E_{sym}$-$L_{sym}$ correlation is estimated to be 0.52-0.53, which is not so different from the one deduced from the $\sim$50 functionals and imposing 
$28<E_{sym}<36$~MeV. The blurring of the $E_{sym}$-$L_{sym}$ correlation shown in Fig.~\ref{fig:EL2} can therefore be considered as realistic of the model dependence of this correlation.
The anticorrelation coefficient for the $K_{sym}$-$(3E_{sym}-L_{sym})$ correlation is estimated to be -0.47-(-0.61), which is also similar to the one deduced from the $\sim$50 functionals and imposing $28\leq E_{sym}\leq 36$~MeV and $40\leq L_{sym} \leq 60$~MeV.
The extremely weak correlation coefficient for the $Q_{sat}$-$K_{sat}$ correlation -- 0.17-(-0.06) -- shown in Fig.~\ref{fig:MM} reflects our conclusions from our previous analysis as well: the $Q_{sat}$ and $K_{sat}$ empirical parameters are very weakly correlated by either the experimental parameter $M_c$ or by the MBPT predictions in symmetric matter below saturation density.

\section{Conclusions and outlooks}

In this paper, we have examined the quality of the correlations among the EoS empirical parameters, coming from the existence of general physics constraints on the EoS, as well as from empirical measurements.
Specifically, we have analyzed the origin and the model dependence of the correlation between $E_{sym}$ and $L_{sym}$, largely observed in the literature, as well as the correlation between $K_{sym}$ and  $3E_{sym}-L_{sym}$, recently proposed~\cite{Mondal2017}, and we have further analyzed the reason of the very weak $Q_{sat}$-$K_{sat}$ correlation.

Within a simple analytical Taylor expansion of the EoS around saturation, we have confirmed that the $E_{sym}$-$L_{sym}$ correlation arises from the empirical knowledge of the symmetry energy at density slightly below saturation, as obtained for example from the IVGDR measurement, and we have estimated its width coming from model dependence. We have found that the main source of uncertainties in this correlation is coming from $K_{sym}$ while the contribution of the higher order empirical parameters ($Q_{sym}$ and $Z_{sym}$) are negligible.
Concerning  the correlation  $K_{sym}$-$(3E_{sym}-L_{sym})$, we have shown that it trivially emerges from the boundary condition on the neutron matter energy density, which explains why it is universally respected. However, when only the functionals corresponding to realistic values of $E_{sym}$ and $L_{sym}$ are retained, the quality of the correlation considerably worsens. 
Finally, we found that $Q_{sat}$ and $K_{sat}$ are weakly correlated, as expected from previous studies~\cite{Khan2012,Margueron2018a}.
These results have been confirmed within a more evolved meta-modeling of the EoS.

We have explained the origin of the dispersion among these correlations from the effect of the high order EoS parameters.
Indeed, while the values of $E_{sym}$ and $L_{sym}$ are relatively close among the different functionals and functional families, high order parameters
such as the isovector incompressibility $K_{sym}$ and the skewness and kurtosis $Q_{sym},Q_{sat},Z_{sym},Z_{sat}$ are largely model dependent.
For phenomenological approaches, this model dependence is mainly due to the small number of free parameters and to the absence of experimental constraints.
The high order empirical parameters are functions of the same model-coefficient as the low order ones, inducing such kind of spurious correlations.
We have shown that the dispersion of the $E_{sym}$-$L_{sym}$ correlation can be nicely understood from the propagation of the uncertainties of $K_{sym}$,
while the $K_{sym}$-$(3E_{sym}-L_{sym})$ correlation is mostly affected  by the uncertainty on $Q_{sat}$ and $Q_{sym}$. 
The weak $Q_{sat}$-$K_{sat}$ correlation induced by either the experimental parameter $M_c$ or the MBPT predictions can be explained from the small contribution of $Q_{sat}$ to the incompressibility below saturation density. The determination of $Q_{sat}$ shall therefore be better constrained by experiments probing matter properties above saturation density, such as for instance heavy-ion collisions.

In conclusion, we have illustrated the complexity in determining the empirical parameters of nuclear matter from correlations between an observable and a single empirical parameter. In the present cases, we have shown the important contribution of the unknown empirical parameter $K_{sym}$ (resp. $Q_{NM}$) on the blurring of the $E_{sym}$-$L_{sym}$ (resp. $K_{sym}$-$(3E_{sym}-L_{sym})$) correlation.
This complexity suggests that in the future, multi-parameter correlation analyses -- satisfying a set of experimental contraints -- shall better be performed to provide better posterior probabilities for the empirical parameters. A meta-modeling, such as the one employed here, is a well adapted tool to perform such statistical analyses.

 \begin{acknowledgments}
This work was partially supported by the IN2P3 Master Project MAC, "NewCompStar" COST Action MP1304, PHAROS COST Action MP16214.
\end{acknowledgments}

\end{document}